# Uncertainty Management in Software Projects: A Case Study in a Public Company

[1]Karina Macedo, [2]Marcelo Marinho, [3]Simone Santos
[1] Federal Institute of Bahia (IFBA) Information Technology (TI) Barreiras, BA, Brazil
Email: karina@ifba.edu.br
[2] Federal Rural University of Pernambuco (UFRPE) Department of Computer Science (DC)
Recife, PE, Brazil, Email: marcelo.marinho@ufrpe.br
[3] Federal University of Pernambuco (UFPE) Informatics Center (CIn) Recife, PE, Brazil
Email: scs@cin.ufpe.br

*Abstract*
*Software development meets the various challenges of rapidly growing markets. To address such challenges projects to design and adopt specific development approaches. However, various project management approaches do not consider the uncertainties that exist in projects. In this paper, we present findings from a case study in which we explore how to apply the Management Uncertainty Software Project (MUPS) approach. We do so by the empirical investigation at a public organization in Brazil. The objective of this study is to contribute to the body of knowledge regarding the potential benefits of MUSP approach. The conclusions of the empirical study will help both researchers and practitioners to understand better which benefits are already being realized in practice, and how they can best be realized.*

## 1. Introduction

With the growing advancement of Information Technology (IT), public organizations structural departments were compelled to deploy new systems to not only match their business needs but also execute operations through the best IT practices with the necessary efficiency and speed. Organizations may use changing processes in the current globalized market to implement these systems [1].

Despite previous studies, various project management approaches do not consider the uncertainties that exist in projects [4]. It is easy to believe that when risks are managed, uncertainties are also managed: these concepts are not identical, as risks are typically quantified regarding probability and the impact of their consequences, but uncertainties are not [5].

This research's initial motivation involves the complexity in implementing an integrated business management system in the public sphere and adopting uncertainty management approach. This implementation revealed the opportunity to investigate adherence to this type of management in an ERP system's educational module within a federal teaching institution (the Unified Public Administration System - UPAS).

This paper presents a case study from the public sector, using Marinho's [11] proposed approach for the management of uncertainties in software projects. This study primarily aimed to coordinate and control an ERP project.

In addition to this introductory section, this paper is structured as follows: Section 2 presents the background; Section 3 presents the adopted research method; Section 4 presents the case study; Section 5 analyses and finally, Section 6 contains conclusions and some directions for future work.

## 2. Background

### 2.1 Uncertainty Management

Johansen et al. [12] provide a process for uncertainty management but do not instruct the project manager and team regarding how to become aware of the early signs of uncertainty, or identifying their associated risks. Martinsuo, Korhonen, and Laine [13] noted how to address uncertainty in program management. Specifically, the authors aspire to understand how portfolio managers handle the threats and opportunities that generate uncertainty. Ramasesh and Browning [14] presented a theoretical framework in which factors and relationships are proposed that increase projects' levels of uncertainty unknown unknowns. Further, Marinho et al. [3] presented a series of publications in the software management field, including an exploratory literature review aiming to identify the basic concepts and primary research sources in uncertainty management. Marinho et al. [6] offered a systematic literature review on managing uncertainty in projects, and action research was conducted using an innovative software project [8]. Semi-structured interviews demonstrated project managers' practical perspectives as well as project management researchers' perspectives [9]. Finally, Marinho et al. [11] denoted an approach to managing software uncertainties.

### 2.2 Uncertainty Management in Software Projects

Marinho [11] presents a theoretical approach to managing uncertainties in software projects, (hereafter referred to as "MUSP"), that is based on six practice sets to guide the software team:
• Characterizing Projects: understanding the best management for the approach to be applied;
• Identification of Uncertainty Sources: identifies the most unfamiliar sources of uncertainty in the project;
• Early Signs Detection: observes the signs of uncertainties perceived in the project;
• Sensemaking: the sense is created to detect signs;
• Risk Management: the phase in which identified risks are managed;
• Unexpected Results: the preparation for, and reaction to, events not anticipated in the project.
Additionally, MUSP presents general guidelines for managers to handle uncertainties and displays a set of proactive techniques, practices, and strategies to reduce or eliminate project uncertainties.

### 3. Research Method

The research was divided into four stages: a research proposal, applied methodology, case study and results analysis. Therefore, the following topics will introduce and detail the steps in this research.

1) Research Proposal: This research study began with an ad hoc literature review [15], with the purpose of studying the elementary concepts that guide this study. In this phase, we also studied systems that could be adapted to conduct this research within the chosen organization. The research problems and their objectives were then defined.

2) Applied Methodology: We used this case study selection phase to study which systems in the chosen organization were in a phase to begin implementation. The case study was designed and planned in stages to provide a better perception of the problem [16].

3) Case Study: MUSP [11] was applied to begin in January 2017, during the case study, with the following phases selected by the project manager: characterization, identification of uncertainty sources, early signs detection and sensemaking.

The data collected from meetings and workshops, used to elaborate upon and apply the steps in the approach, were then transcribed and analyzed. These meetings consisted of conversations with project participants as well as on-site observations. After this analysis, the collected documentary evidence and the researchers' observations were consolidated, and the results reported.

4) Results Analysis: Data analysis aims to derive conclusions from collected data both clearly and systematically by maintaining a consistent chain of evidence [16]. Further, Merriam [17] posited that data collection and analysis should be a simultaneous process in qualitative research, and not sequential. Specifically, while the collection and analysis of data is a resourceful and dynamic process, this does not mean that the analysis ends when all data is collected, but that this analysis becomes more intense as the study progresses. Thus, the data were collected, coded and analyzed throughout the research process.

## 4. Case study findings and analysis

We applied this study Department of Information Systems (DIS) at the federal institute. Intensity sampling, which targets a larger number of interview participants with different responsibilities within the same unit of analysis, was employed to obtain richness and depth in the study. Perspectives from participants with different responsibilities were obtained to triangulate the data. Responsibilities included: developers, testers, project management, and corporate-level executives.

Further, we applied this study in UPAS Software Team, which was chosen to meet the institution's most urgent demands. The first module to be deployed was UPAS-EDU, the academic module, due to a substantial collective interest in the unified use of school records.

### 4.1. MUSP Application in UPAS-EDU

We conducted some evaluations to assess how the organization perceives the uncertainties arising from a software project development. This led to a collection of the organization's perceptions and actions regarding the uncertainties in software projects.

The study was supported with documentary sources, such as publicly available white papers, technical reports, case studies and web hosted marketing materials. On-site visits to secure work environments enabled first-hand observation of working practices and workplace environments. Teams coordination meetings were observed.

However, the primary data collection technique employed in the study was face-to-face interviews conducted with practitioners performed in January and November 2017. The following subsections will present each phase's results.

*1) Project Characterisation*: Table 1 illustrates a data set identified in project characterization phases, such as the chosen management methodology, a stakeholder analysis, and project definition criteria.

**Table 1.** Project characterization.

| Types of Project Management | Traditional Project Management was chosen because the project objectives are known, and the project team can define the same solutions. | | | |
|---|---|---|---|---|
| **Stakeholder Analysis** | **High Power X High Interest** | **High Power X Low Interest** | **Low Power X High Interest** | **Low Power x Low Interest** |
|  | High management of the Institute: Rector, Teaching Pro-rector, GDIT and campuses of Directors. | The other Pro-rectories. | Academic records sector and IT campuses. | Other administrative sectors of the Rectory and campuses. |
| **Success Definition Criteria** | **Customer Satisfaction and Impact** | **Motivation and Team Impact** | **Efficiency and Efficacy** | **Prepare the Future** |
|  | • The synchronization of institutional data, as well as updated reports of school records between campuses; • Maintain the flow and dynamics of information accessed at any level in the institution; • Uniform academic system in use throughout the institution. | • Measuring learning, enthusiasm, motivation, and team loyalty; • Professional satisfaction; • Acquiring new experiences and learning; • Elaborate courses and training. | • The Entire process is well-managed. | • Prepare the institution to adhere to the other UPAS modules; • Ensure team's expertise for the deploy of subsequent modules; • Promote efficient and effective data manipulation across all campuses and facilitate communication among all stakeholders. |

*2) Uncertainty Sources Identification:* MUSP proposes strategies in this step to clarify uncertainty sources. We applied Diagram of Cause and Effect and Building Scenarios that they have revealed: estimation errors, lack of process control and synchronization in academic standards; lack of expertise in the migration and development platform; lack of financial resources; high-pressure management; lack of commitment among members and lack of expertise in the development and migration platform. Thus, the team defined to the degree source of uncertainty. The higher the value, the more secure the team felt about the sources of uncertainty. Table 2 describes this.

**Table 2.** Uncertainties raised by the team.

| Uncertainty Sources | Uncertainties | Values |
|---|---|---|
| **Technological** | U1 - Expertise on the extraction and data migration platform | 3 |
| | U2 - Absence of a formal tool or method for time, scope, risk and quality management | 2 |
| | U3 - System platform expertise | 1 |
| **Socio-Human** | U4 - Relationships between senior managers and end users | 5 |
| | U5 - the Insufficient team for system demands | 4 |
| | U6 - Manager and team relationship | 4 |
| | U7 - Sectors of the business area seldom involved in the project | 3 |
| | U8 - The development teams internal relationship | 3 |
| | U9 - Team motivation | 3 |
| | U10 - Cultural, political, or religious values, beliefs, and experiences and their interference in project management | 3 |
| | U11 - Resilience to overcome difficulties | 2 |
| | U12 - Members solitary tasks without effective management | 2 |
| | U13 - Language knowledge restricted to project personnel | 2 |
| **Environmental** | U14 - High-pressure management for fast delivery | 4 |
| | U15 - Lack of financial resources | 3 |
| | U16 - Emergency installations of other systems concurrently | 3 |

*3) Detection of early signs:* Marinho [10] proposes the use of mindfulness attributes in this phase, to be applied during the project's entire development cycle. After presenting to the team and manager all the guidelines prescribed in the approach regarding these attributes' application. The team believes that the manager must be more present in the project by coordinating cycle planning and interfering in activities that require adjustments. However, the manager sequentially believes that the team must learn to develop without a constant managerial presence, and recognize their faults and inconsistent actions in the project. Nonetheless, the team and manager could identify some early signs of uncertainty presented in Table 3.

**Table 3.** Early signs raised by the team

| Uncertainty Label | Early Signs of Uncertainty |
|---|---|
| U14 | ES1 - Placing and disseminating the system in production only to have an "accomplished" status, without even measuring or performing the procedures essential for initial use. |
| U7 | ES2 - Top management's inflexibility in understanding the implementation in segments. |
| U1, U2, U3 | ES3 - Programming platform expertise restricted to one team member |
| U1, U3 | ES4 - Knowledge and crucial project activities concentrated on a single team member. |
| U5, U16 | ES5 - Team dissolved and fragmented to handle other projects |
| U6 | ES6 - Manager involved in multiple projects; this impedes monitoring of all the signals as well as complete awareness of the project |
| U8, U12 | ES7 - A lack of synergy between old and new team members |
| U7 | ES8 - Need for more support from top management to apply the system and all its components in all its artifacts |
| U8, U9 | ES9 - The team performs solitary tasks based on each member's own experiences, and these are often not shared |
| U9, U11 | ES10 - Insecurity of some members regarding the system, caused by the frustrations with the old ERP System, in which they had to redo and even lose many jobs |
| U10, U13 | ES11 - Misalignment between cultural and political beliefs and values in the project management's progress |
| U14, U5 | ES12 - Number of staff members insufficient to meet development, migration and testing demands in UPAS-EDU |
| U15 | ES13 - The organization's lack of financial resources |

It can be observed that provoking the team through approaches to these practices has fostered nuisances, necessary for them to realize what is being done and how it should be done.

*4) Sensemaking:* After discovering all the early signs, it is now the responsibility of the team minimize or negate these signs. The sensemaking phase translates each signs trajectory into a risk. The team interprets this sign, and the manager provides input regarding their beliefs and perceptions; these signs are then shared and interpreted, illustrating the project's risks.

The activities in this step were conducted following the sensemaking process adopted by MUSP, as follows: (i): The project manager was defined as the sensemaker, and began to interpret the signs by considering several factors that could correlate with the projects; (ii): The team was encouraged to translate all perceived signs to convert this into actions that could be understood by the entire team; (iii): The manager was responsible for considering each team member's experience during stage ii and (iv): All meanings for each early sign were clearly conveyed to the team, which then discussed the necessary steps to be taken.

Table 4. Some risks found from early signs

| Early Signs Label | Risks |
|---|---|
| ES3, ES4, ES5, ES7, ES9, ES10, ES12 | R1 - Delay in the data migration process |
| ES3, ES4, ES5, ES7, ES9, ES10, ES12 | R2 - Delays in the system's execution and delivery |
| ES3, ES4, ES13 | R5 - Difficulties in accessing legacy systems on an adequate platform |
| ES2, ES3, ES8, ES9, ES13 | R7 - Excessive dependence on legacy systems |
| ES2, ES3, ES6, ES8, ES11 | R8 - Distorted evaluation of the system's effectiveness due to ignorance of the institution's academic norms and processes |
| ES2, ES6, ES8, ES11 | R9 - A lack of involvement from the main business sector (pro-rector of teaching) in the project |
| ES2, ES3, ES4, ES5, ES9, ES12 | R11 - The implementation team's lack of breadth to account for demands |
| ES3, ES9, ES12 | R15 - A rework in the system code |
| ES3, ES4, ES7, ES9 | R16 - Not sharing failures and errors |
| ES2, ES6, ES10 | R19 - Frustrations of key stakeholders expectations due to a lack of project management |
| ES6, ES8, ES10 | R20 - Boycotting the new system |
| ES6, ES7 | R25 - A lack of knowledge acquired about the project, such as its tools and processes |
| ES6 | R26 - End users' lack of comprehension about the deployment processes |
| ES2, ES6, ES7, ES10, ES11 | R27 - A mismatch of expectations between staff deployment and management |
| ES3, ES7 | R28 - A lack of standardization in team members' work processes |
| ES6, ES8 | R29 - Desynchronisation of the academic processes across the institute's campuses |
| ES3, ES4, ES5 | R31 - Technical team restructuring |
| ES3, ES5, ES6 | R32 - Loss of a team member who holds knowledge of the system and its adaptations |
| ES3 | R33 - Artefacts and procedures not by academic standards |
| ES5, ES6, ES7, ES9, ES11 | R36 - Unnecessary energy expenditures because the project lacks all artifacts and methodologies for application |
| ES2, ES5, ES6, ES8, ES11 | R37 - Manager responsibilities that are inconsistent with their performance in the project, and that generate disbelief, misalignment, and demotivation in the team |
| ES3, ES6, ES8 | R42 - Errors in interpretation and works performed without consonance with the fundamental needs |
| ES1 | R46 - A key requirement passed by an interested party is neglected or overlooked |
| ES5, ES6, ES8 | R47 - Cancellation or termination of the project |
| ES1, ES6 | R48 - The absence of project documentation |

After the abstraction of the early signs of uncertainty found in UPAS-EDU, some project risks were discovered during the sensemaking phase. Table 4 shows some risks found from the early signs.

## 5. Discussion

The application of the MUSP has revealed many difficulties in its execution. The manager and team were so focused on implementing the UPAS-EDU module that many considerations in this approach would have been neglected if the researchers had not interfered and assisted them in its implementation.

The characterization phase was easily executed, but there were some doubts regarding the types of stakeholders; each was consensually designated, and it was possible to establish a project management methodology. The identification phase for the sources of uncertainty was the most time-consuming because the team was not accustomed to creating diagrams, scenarios. This phase revealed that the main sources of uncertainty in the project were technological and socio-human. Additionally, the team's need to build a relationship of trust and continuous learning was sharply realized, and thus, the steps for UPAS-EDU were efficiently executed.

Mindfulness practices were applied during project development in the early signs detection phase. It was demonstrated how the team and manager had neglected several signs of uncertainty. The team realized the need to adopt mindfulness practices to both capture signals and improve project management. The team realized at the end of this phase that the questions addressed in mindfulness and recommendations should be included in the UPAS-EDU team's daily processes, and extended to other institutional projects.

Thus, these signs were interpreted in consensus in the subsequent sensemaking phase and transformed into risks. Additionally, sensemaking meetings allowed the research team to meet with the project manager and present points of improvement while conducting sensemaking activities.

The risk extraction stage was then conducted in corroboration with prior analyses with the same team and project manager. This also contributes to the conclusions regarding the primary sources of uncertainty that deserve more attention in the UPAS-EDU project.

If uncertainty management received the full support of top management and project managers, it could highlight and respond to most inconsistent practices in this project.

## 6. Conclusion

MUSP helped reveal factors and measures to improve UPAS-EDU project. Further, some improvements can be observed in the organization: an increased awareness of uncertainties within the institutions projects; motivation regarding the urgency in adopting an application methodology in the project; initial actions in response to initially discovered signs of alarm; guiding strategies to use in detecting possible sources of uncertainty; and the origin of input for analysis and future project risk management.

As with all academic research, this study has limitations that must be considered, as follows: The number of study participants was limited due to the size of the project team, and the results were restricted to a public institution in the implementation of an ERP system under the uncertainty management view. Further, as the phases in the MUSP were not fully considered, the institution's participation in this framework was adapted to its reality and the chosen phases of execution.

Despite faced with such limitations, this research was guided with all the necessary methodological rigor to mitigate threats to validity. This study aimed to demonstrate a reliable result, allowing for some contributions to the community as well as its representation in future studies.

Although the initially proposed objectives were reached, other issues complementary to the application of the uncertainty management approach appeared during this study, which was not investigated. These issues deserve to be further developed in new research: (i) application in public organizations with mature project management practices, to identify the primary sources of uncertainty and act on them; and (ii) application in private organizations, which adopt uncertainty management as an ally to their project management. This allows for a comparative evaluation between implementation with and without the use of this approach.